\def\w0{\omega_0}

\documentstyle[prb,aps,epsfig,multicol]{revtex}

\begin{document}

\title{Stochastic resonance for two competing species in the presence of colored noise}
\author{\large D. Valenti, A. Fiasconaro, B. Spagnolo\\
{\em \normalsize Dipartimento di Fisica e Tecnologie Relative dell'Universit\`a di Palermo}\\
{\em \normalsize Istituto Nazionale di Fisica della Materia, Unit\`a di Palermo}\\
{\em \normalsize Viale delle Scienze, I-90128 Palermo, Italy}}
%
\maketitle
\begin{abstract}
We study the role of multiplicative colored noise for different
values of the correlation time $\tau_c$ in the dynamics of two
competing species, described by generalized Lotka-Volterra
equations. The multiplicative colored noise models the interaction
between the species and the environment. The interaction parameter
between the species is a random process which obeys a stochastic
differential equation with a generalized bistable potential in the
presence of a periodic driving term, which accounts for the
environment temperature variation. The bistable potential is
useful to describe the coexistence and exclusion dynamical regimes
of the ecosystem. Noise-induced periodic oscillations of the
species concentrations and stochastic resonance phenomenon appear
due to the presence of the multiplicative noise. We find that for
low values of the correlation time $\tau_c$ the response of the
system coincides with that obtained with multiplicative white
noise. For higher values of $\tau_c$ the coherent response of the
system and the maximum of the signal-to-noise ratio, signature of
the stochastic resonance phenomenon, are shifted towards higher
values of the noise intensity.
\end{abstract}
%
%
\section{Introduction}\indent
Recently the constructive role played by the noise in population
dynamics has been the subject of intense theoretical
investigations~ \cite{Ciu,Wio,Sch,Sta,Spa1}. The noise, through
its interaction with the nonlinearity of the ecosystems, can give
rise to counterintuitive phenomena like stochastic
resonance~\cite{Gam} (SR), noise enhanced stability~\cite{Agu}
(NES), noise delayed extinction ~\cite{Spa2}, etc...

In this work we focus on the role of realistic noise in the
dynamics of two competing species, and specifically on the SR
phenomenon in population dynamics in the presence of exponentially
correlated noise. When the time scale of random fluctuations is
larger than the characteristic time scale of the ecosystem the
external noise cannot be considered white noise. This noise has no
time scale because is $\delta-correlated$. In real ecosystems the
external random perturbations, due to interaction with the
environment, are correlated with a finite correlation time. A
strongly correlated noise for example emerges as the result of a
coarse graining over a hidden set of slow variables~\cite{Gam}.
The dynamics of our ecosystem is described by generalized
Lotka-Volterra equations with a random interaction parameter
between two competing species in the presence of a multiplicative
colored noise. In the absence of noise we have two stable states,
corresponding to dynamical regimes of coexistence and exclusion.
Real ecosystem however are subjected to noisy environment and to
geological periodical forcing, like seasonal temperature. Because
of these random and periodical forcing we describe the two
dynamical regimes of coexistence and exclusion with a bistable
potential for the interaction parameter, tilted periodically by an
environment temperature driving term.

We find that the noise is responsible for the generation of
quasi-periodic temporal oscillations and the enhancement of the
response of the system to a driving force producing stochastic
resonance~\cite{Spa2}. By analyzing the effect of the colored
noise we find that for low values of the correlation time $\tau_c$
the response of the system coincides with that obtained with
multiplicative white noise. For higher values of $\tau_c$ the
coherent response of the system and the maximum of the
signal-to-noise ratio (SNR), which are signature of the SR
phenomenon, are shifted towards higher values of the noise
intensity. These results are in agreement with previous
theoretical and experimental investigations of SR phenomenon in
dynamical systems in the presence of colored
noise~\cite{Gam,Man,Han}. However in previous studies the colored
noise was additive, while here we have two different sources of
noise and only one of them is colored.
\section{The model}

Time evolution of two competing species is obtained within the
formalism of the Lotka-Volterra equations~\cite{Lot} in the
presence of a multiplicative noise
\begin{eqnarray}
\frac{dx}{dt}=\mu_1\thinspace x\thinspace(\alpha_1-x-\beta_1(t) y)+x\thinspace\zeta_x(t)\\
\frac{dy}{dt}=\mu_2\thinspace y\thinspace(\alpha_2-y-\beta_2(t)
x)+y\thinspace\zeta_y(t),
 \label{LotVol}
\end{eqnarray}
where $\zeta_i(t)$ $(i=x,y)$ are colored noises given by the
archetypal source for colored noise, i. e. exponentially
correlated processes given by Ornstein-Uhlenbeck
process~\cite{Gar}
\begin{equation}
\frac{d\zeta_i}{dt}=-\frac{1}{\tau_c}\zeta_i + \frac{1}{\tau_c}
\xi_i(t) \qquad (i=x,y)
\label{colored_noise}
\end{equation}
and $\xi_i(t)$ $(i=x,y)$ are Gaussian white noises within the Ito
scheme with zero mean and correlation function $\langle
\xi_i(t)\xi_j(t')\rangle = 2\sigma \delta(t-t')\delta_{ij}$. The
correlation function of the processes of
Eq.($\ref{colored_noise}$) is
\begin{equation}
\langle \zeta_i(t)\zeta_j(t')\rangle = \frac{\sigma}{\tau_c}
e^{-|t-t'|/\tau_c} \delta_{ij}
 \label{correlation function}
\end{equation}
and gives $2\delta(t-t')\delta_{ij}$ in the limit $\tau_c
\rightarrow 0$. We can distinguish two different regimes for the
correlation time of the noise: (a) $dt < \tau_c < T_o$, i. e. a
correlation time greater then the integration step $dt$ and less
then the period of the deterministic driving force $T_o$; (b)
$\tau_c > T_o$.

The time series for the two populations are obtained setting
$\alpha_1=\alpha_2=\alpha $, $\beta_1(t)=\beta_2(t)=\beta(t)$, and
$\xi_x(t)=\xi_y(t)=\xi(t)$. For $\beta < 1 $ both species
survives, that is a coexistence regime takes place, while for
$\beta > 1 $ an exclusion regime is established and one of the two
species extinguishes after a certain time. Coexistence and
exclusion of one of the two species correspond to stable states of
the Lotka-Volterra's deterministic model~\cite{Baz}. This aspect
together with the assumption of a strong noisy interaction between
species and environment suggests that the interaction parameter
$\beta(t)$ can be described as a stochastic process which obeys
the following stochastic differential equation
\begin{equation}
\frac{d\beta(t)}{dt} = -\frac{dU(\beta,t)}{d\beta} +
\xi_{\beta}(t),
 \label{beta_eq}
\end{equation}
where $U(\beta,t)$

\begin{equation}
U(\beta,t) =
h(\beta-(1+\rho))^4/\eta^4-2h(\beta-(1+\rho))^2/\eta^2 -\beta
\gamma cos(\omega_0 t) ,
\label{U(beta,t)}
\end{equation}
is a bistable potential (see Fig.1), with the two stable states of
coexistence and exclusion, modulated by the periodic term $\gamma
cos(\omega_0 t)$, which takes into account for the environment
temperature oscillations. Here $h$ is the height of the potential
barrier.
\begin{figure}[htbp]
\begin{center}
\includegraphics[width=9cm]{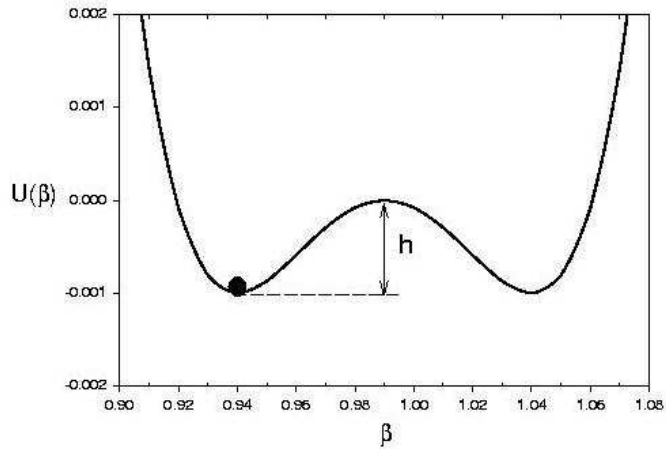}
\end{center}
\caption{ \small \emph{The bistable potential $U(\beta,t)$ of the
interaction parameter $\beta(t)$ at $\omega_o t = \pi/2$. The
potential $U(\beta,t)$ is centered on $\beta=0.99$. The parameters
of the potential are
 $h = 6.25 \cdot 10^{-3}$, $\eta=0.05$, $\rho = -0.01$.}\bigskip}
\label{potential}
\end{figure}
In Eq.~(\ref{beta_eq}) $\xi_{\beta}(t)$ is a Gaussian white noise
with the usual statistical properties: $\langle
\xi_{\beta}(t)\rangle=0$ and $\langle
\xi_{\beta}(t)\xi_{\beta}(t')\rangle = \sigma\delta(t-t')$. We
choose as initial conditions for two competing species a
coexistence regime ($\beta(0) < 1$) on the left minimum of the
potential $U(\beta,t)$ at $\omega_0 t = \pi/2$.
\subsection{Stochastic resonance}
\begin{figure}[htbp]
\begin{center}
\includegraphics[width=10.5cm]{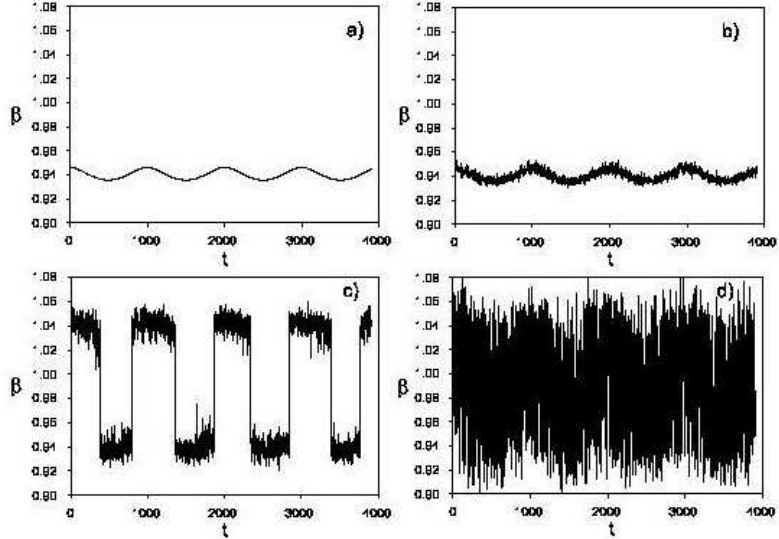}
\end{center}
\caption{ \small \emph{Time evolution of the interaction
parameter, with initial value $\beta(0)=0.94$, for different
values of the additive noise intensity $\sigma_\beta$. (a)
$\sigma_\beta=0$; (b) $\sigma_\beta=1.78\cdot 10^{-4}$; (c)
$\sigma_\beta=1.78\cdot 10^{-3}$; (d) $\sigma_\beta=1.78\cdot
10^{-2}$. The values of the other parameters are $\gamma=10^{-1}$,
$\omega_0/(2\pi)=10^{-3}$.}\bigskip} \label{beta_series}
\end{figure}

The deterministic dynamics of the species strongly depends on the
value of the interaction parameter $\beta(t)$. We first analyze
therefore the time evolution of $\beta(t)$ for different levels of
the additive noise $\sigma_\beta$~\cite{Val}. In the absence of
noise ($\sigma_\beta=0$) we have a periodical behaviour of
$\beta(t)$ in the coexistence regime (see Fig.\ref{beta_series}a).
For low noise intensity with respect to the height of the
potential barrier ($\sigma_\beta \ll h$) we obtain a
quasi-deterministic trajectory of $\beta(t)$, slightly perturbed
by the noise (Fig.\ref{beta_series}b). When the noise intensity is
comparable with the barrier height ($\sigma_\beta \simeq h $), the
typical picture of stochastic resonance phenomenon is obtained:
the periodical behavior of the interaction parameter jumps between
the two values $\beta=0.94<1$ and $\beta=1.04>1$, which
characterize the coexistence and the exclusion dynamical regimes
(see Fig.(\ref{beta_series}c). A further increase of the noise
intensity produces a loss of coherence and the dynamical behavior
of $\beta(t)$ is strongly controlled by the noise
(Fig.\ref{beta_series}d).
The optimum coherent time behaviour of $\beta(t)$
(Fig.(\ref{beta_series}c) is obtained from the statistical
synchronization relation

\begin{equation}
\tau_k=T_0/2 \label{SR_sync}
\end{equation}
typical of the SR phenomenon. Here $\tau_k$ is the Kramers time

\begin{equation}
\tau_k = \frac{2 \pi}{\sqrt{\vert U''(0.99) \vert U''(0.94)}}
\exp{[2 h/\sigma_\beta]},
\label{kramers}
\end{equation}
$T_0=2\pi/\omega_0$ is the period of the driving force,
$U''(0.99)$ and $U''(0.94)$ are the second derivative respectively
calculated in the unstable and stable states of the potential. For
$T_o = 10^3$ we have $\sigma_\beta=1.78\cdot 10^{-3}$, the value
of the noise intensity of Fig.\ref{beta_series}c.

\begin{figure}[htbp]
\begin{center}
\includegraphics[width=10cm]{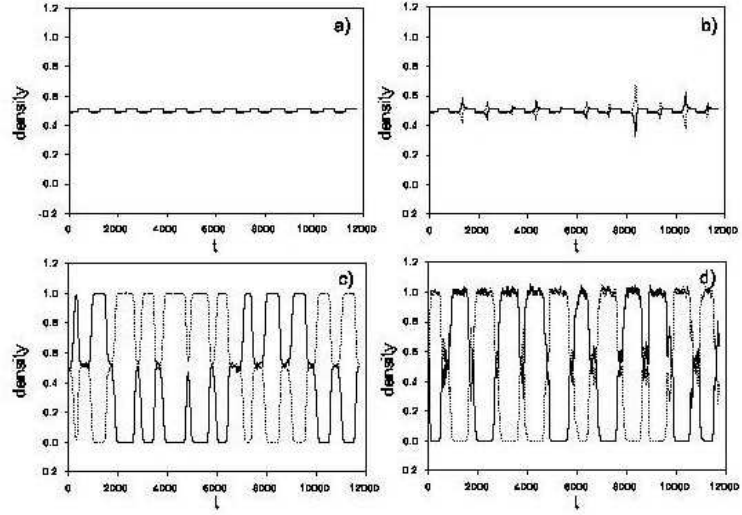}
\end{center}
\caption{ \small \emph{Time evolution of both populations at
different levels of the multiplicative noise for $\tau_c=20$: (a)
$\sigma=0$; (b) $\sigma=10^{-12}$; (c) $\sigma=10^{-4}$; (d)
$\sigma=10^{-2}$.The values of the parameters are $\mu = 1$,
$\alpha=1$, $\gamma = 10^{-1}$, $\w0/(2\pi) = 10^{-3}$. The
intensity of the additive noise is fixed at the value
$\sigma_\beta=1.78 \cdot 10^{-3}$. The initial values are: for the
two species $x(0)=y(0)=1$, for the additive (white) noise
$\beta(0)=0.94$, for the multiplicative (colored) noise
$\zeta_1(0)=\zeta_2(0)=0$.}\bigskip}
\label{time_series_3}
\end{figure}

\begin{figure}[htbp]
\begin{center}
\includegraphics[width=10cm]{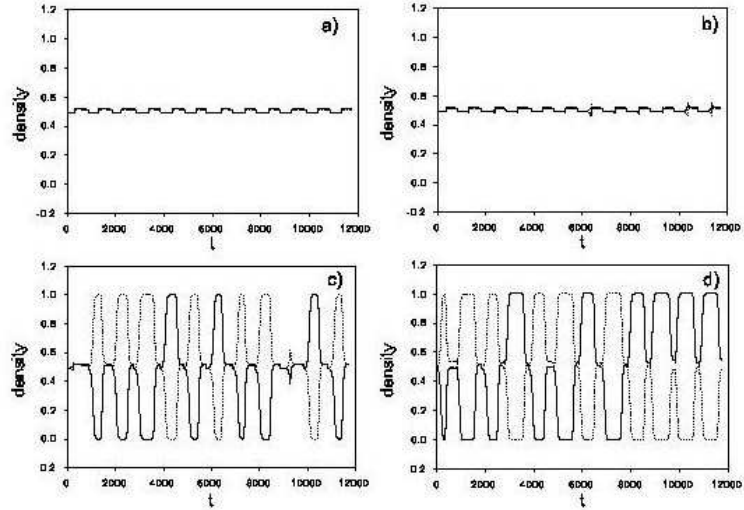}
\end{center}
\caption{ \small \emph{Time evolution of both populations at
different levels of the multiplicative noise for
$\tau_c=2\cdot10^3$: (a) $\sigma=0$; (b) $\sigma=10^{-12}$; (c)
$\sigma=10^{-4}$; (d) $\sigma=10^{-2}$. The values of the
parameters, the intensity of the additive noise and the initial
conditions are the same of
Fig.\ref{time_series_3}.}\bigskip}
\label{time_series_4}
\end{figure}

\begin{figure}[htbp]
\begin{center}
\includegraphics[width=10cm]{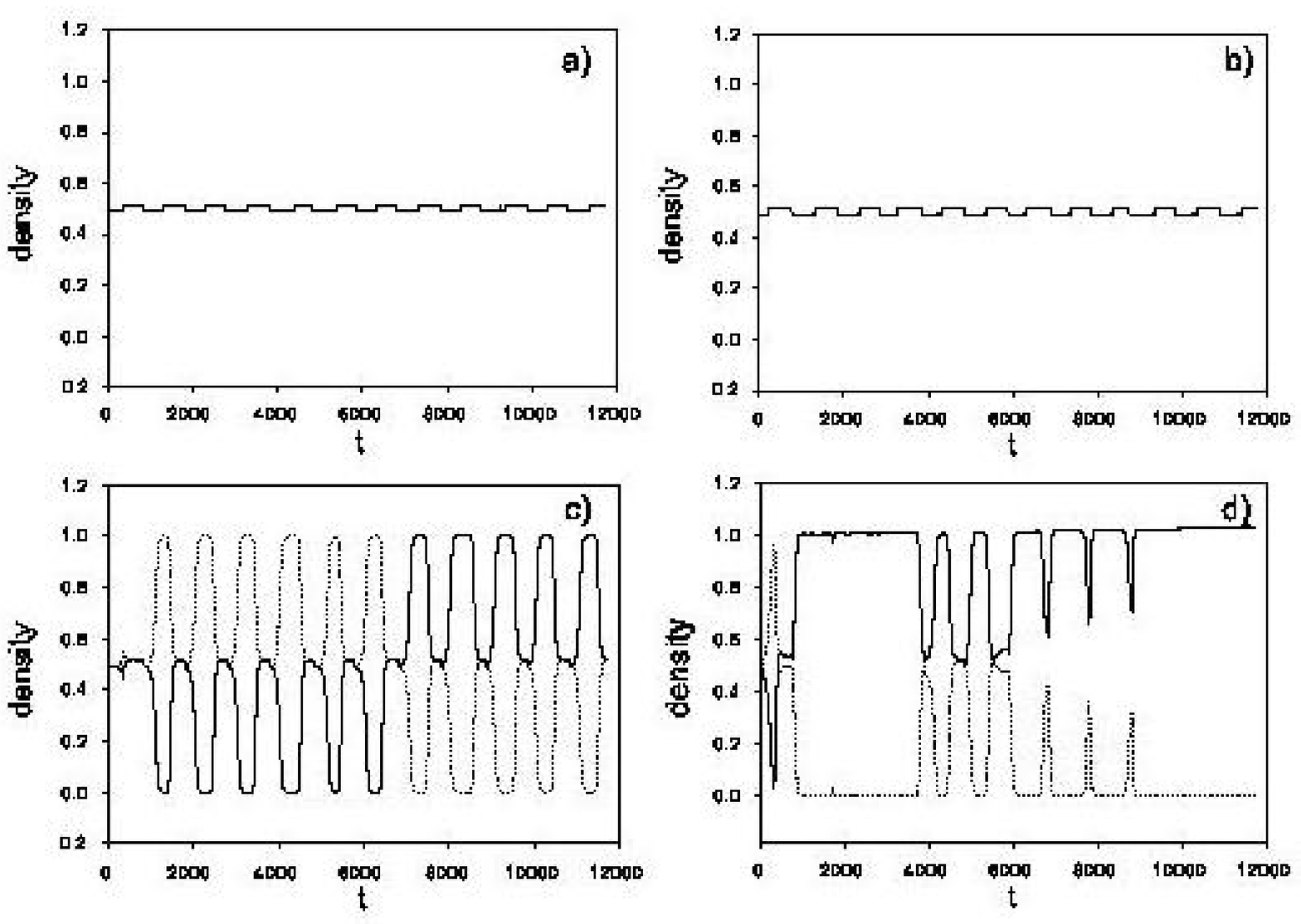}
\end{center}
\caption{ \small \emph{Time evolution of both populations at
different levels of the multiplicative noise for
$\tau_c=2\cdot10^6$: (a) $\sigma=0$; (b) $\sigma=10^{-12}$; (c)
$\sigma=10^{1}$; (d) $\sigma=10^{5}$. The values of the
parameters, the intensity of the additive noise and the initial
conditions are the same of Fig.\ref{time_series_3}.}\bigskip}
\label{time_series_5}
\end{figure}

\begin{figure}[htbp]
\begin{center}
\includegraphics[width=10cm]{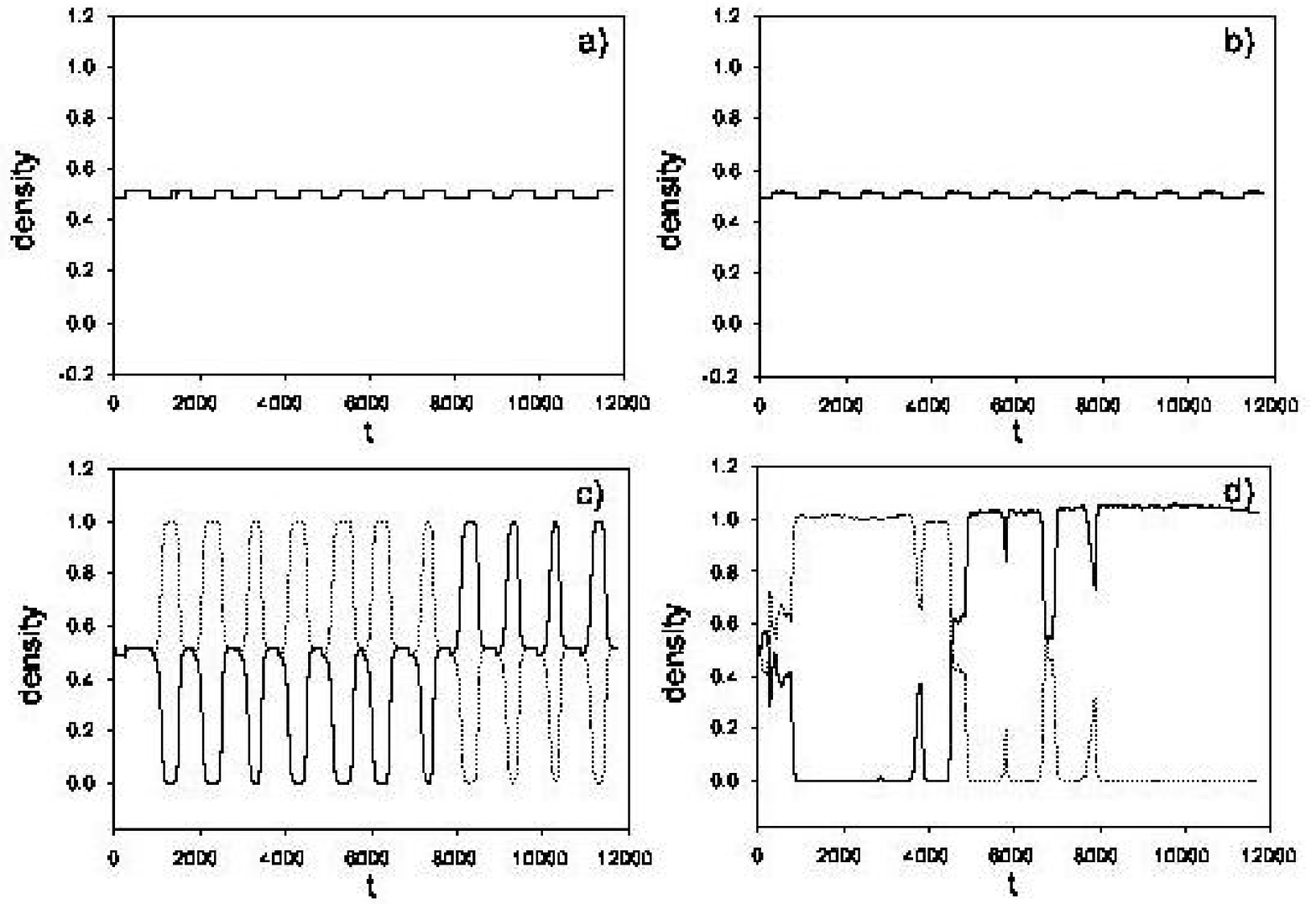}
\end{center}
\caption{ \small \emph{Time evolution of both populations at
different levels of the multiplicative noise for
$\tau_c=2\cdot10^9$: (a) $\sigma=0$; (b) $\sigma=10^{-12}$; (c)
$\sigma=10^{7}$; (d) $\sigma=10^{12}$. The values of the
parameters, the intensity of the additive noise and the initial
conditions are the same of Fig.\ref{time_series_3}.}\bigskip}
\label{time_series_6}
\end{figure}

Now we analyze the dynamics of the two species by fixing the
additive noise intensity at the value $\sigma_\beta=1.78\cdot
10^{-3}$, corresponding to a competition regime between the two
species periodically switched from coexistence to exclusion, and
by varying the intensity of the multiplicative colored noise. We
obtain the temporal series of the two species for different values
of both the multiplicative noise intensity
$\sigma=\sigma_x=\sigma_y$ and the correlation time $\tau_c$. We
do simulations in both regimes for the correlation time of the
noise: (a) $dt < \tau_c < T_o$, and (b) $\tau_c > T_o$. In the
weak correlated noise regime (a) no relevant modifications occur
in the temporal series of the two species densities. The time
evolution of the two species show an anticorrelated behavior with
quasiperiodical oscillations (see Figs.
\ref{time_series_3}c,\ref{time_series_4}c) with a random inversion
of the population that predominates over the other one, as in the
white noise case ~\cite{Val}. In
Figs.~\ref{time_series_3},\ref{time_series_4} we report these time
behaviours at $\sigma=0$, $\sigma=10^{-12}$, $\sigma=10^{-4}$,
$\sigma=10^{-2}$ for two different values of the correlation time
$\tau_c = 20, 2\cdot10^{3}$. For $0<\tau_c<T_o=10^3$ we obtain the
same behaviour as in the presence of multiplicative white noise.
For $\tau_c \simeq T_o$ some modifications occur, particularly for
$\tau_c = 2\cdot10^{3}$ the temporal series of the two species
densities show the anticorrelated behavior with quasiperiodical
oscillations up to $\sigma=10^{-2}$ (see
Fig.~\ref{time_series_4}d), i. e. a delay in the coherent output
of our ecosystem. This delay will manifest itself in the behaviour
of the signal-to-noise ratio (SNR) as a function of the
multiplicative noise intensity.
In the strong correlated noise regime (b) a relevant delay of the
coherent time behaviour of the two species is observed. The
maximum of SNR is shifted towards higher values of the
multiplicative noise intensity. This shift in a Log-Log scale
grows faster than a linear function of the correlation time
$\tau_c$. In Fig.~\ref{time_series_5},\ref{time_series_6} we
report the temporal series for $\tau_c=2\cdot10^6,2\cdot10^9$
obtained for different values of the multiplicative noise
intensity.
Comparing the temporal series of
Figs.~\ref{time_series_3},\ref{time_series_4} and
Figs.~\ref{time_series_5},\ref{time_series_6} we note that
increasing the values of $\tau_c$ the noise-induced effects are
observed for higher levels of noise intensity. The coexistence
regime and the correlated oscillations of both populations persist
for a wider range of multiplicative noise intensities. The
anticorrelated behavior with quasiperiodical oscillations appear
with very high noise intensity as the correlation time value of
the multiplicative noise is strong enough. The lost of coherence
in the time behaviours of the two species is observed at very high
intensities of the multiplicative noise. Because of the high
values of the multiplicative noise, one population extinguishes
and the other one survives at a constant density after a transient
dynamics (see Figs.~\ref{time_series_5}d,\ref{time_series_6}d).
This dynamical behaviour is typical of an ecosystem in the
presence of an absorbing barrier \cite{Ciu}.

The periodicity of the noise-induced oscillations in the time
behavior of the population densities shown in
Figs.~\ref{time_series_3}-\ref{time_series_6} is the same of the
driving periodic term of Eq.(\ref{U(beta,t)}).
\begin{figure}[htbp]
\begin{center}
\includegraphics[width=14cm]{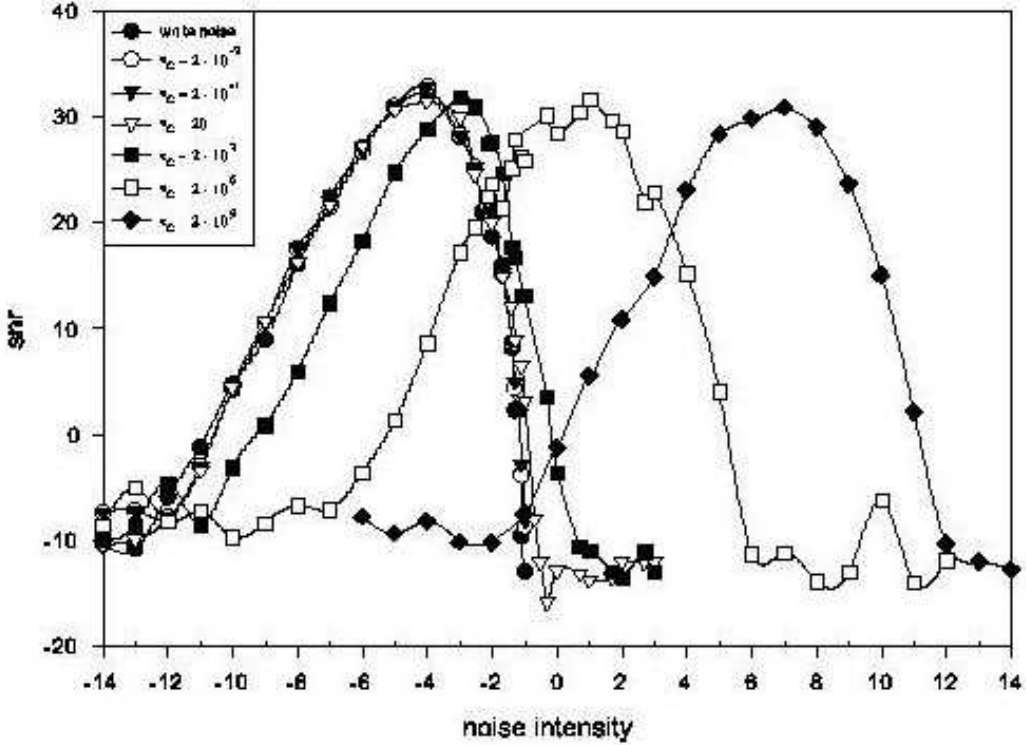}
\end{center}
\caption{ \small \emph{Log-Log plot of SNR as a function of noise
intensity. The SNR is obtained for six different values of the
correlation time: $\tau_c=2\cdot10^{-2}$, $\tau_c=2\cdot10^{-1}$;
$\tau_c=20$; $\tau_c=2\cdot10^3$, $\tau_c=2\cdot10^6$;
$\tau_c=2\cdot10^9$. Moreover the signal-noise ratio for white
Gaussian noise is reported. The SNR corresponds to the squared
difference of population densities $(x - y)^2$. The values of the
parameters and the initial condirtions are the same of those used
to obtain the temporal series.}\bigskip}
\label{snr}
\end{figure}
This is the signature of stochastic resonance (SR)phenomenon. In
order to manifest the presence of SR we analyze the squared
difference of population densities $(x-y)^2$ for different values
of $\tau_c$. In Fig.~\ref{snr} the SNRs of this quantity are shown
for
$\tau_c=0,2\cdot10^{-2},2\cdot10^{-1},20,2\cdot10^3,2\cdot10^6,2\cdot10^9$
 as a function of the multiplicative noise
intensity $\sigma$, by fixing the additive noise intensity at
$\sigma_\beta=1.78 \cdot 10^{-3}$. We calculate the SNR of
$(x-y)^2$ because of the random periodical inversion of population
densities. We note that dynamics of $(x-y)$ is mainly affected by
the multiplicative noise, as we can see from Eqs.(\ref{LotVol}).
The SNRs have been obtained by performing $190$ realizations of
Eqs.~(\ref{LotVol}). In each graph of Fig.~\ref{snr} a maximum
appears, whose position depends on the values of $\tau_c$, i.e.
the most coherent response of the system is connected with both
the intensity and the correlation time of the multiplicative
noise. We see clearly the two dynamical regimes of weak correlated
noise (a), with the first four values of $\tau_c$, and of strong
correlated noise (b) with the last three values of $\tau_c$. In
this regime the maximum of SNR is shifted towards higher values of
multiplicative noise intensity as in previous theoretical and
experimental studies~\cite{Gam,Man,Han}. However some differences
occur. Previous studies on the effect of colored noise on SR
phenomenon showed that by increasing $\tau_c$ the peak of the SNR
shifts towards higher values of the noise amplitude and the
maximum decreases with a broadening of the entire curve. The shift
of the SR peak to larger noise intensities is due to the fact that
colored noise suppresses exponentially the hopping rate with
increasing noise color. In our model the colored noise is
introduced in the multiplicative noise and not in the additive one
as in usual bistable dynamical systems. The SR in the dynamics of
interaction parameter $\beta$ induces SR phenomenon in the
dynamics of two competing populations~\cite{Val}. Our hopping rate
in the first SR is not affected by the \emph{"color"} of the
multiplicative noise. However this noise produces the coherent
response of the ecosystem and therefore produces the shift of the
SNR peak. A more detailed study on the effect of colored noise on
two noise sources used in our model will be the subject of future
investigation.
\section{Conclusions}
We report a study on the role of the colored noise in the dynamics
of two competing species. In our model we consider two Gaussian
noise sources: a multiplicative colored noise and an additive
white noise, which produces a random interaction parameter between
the species. The additive noise controls the switching between the
coexistence and the exclusion dynamical regimes, the
multiplicative noise is responsible for periodical oscillations of
the two species densities, whose amplitude and coherence depend on
the value of the correlation time $\tau_c$. When the interaction
parameter is in SR regime a coherent time behavior of two species
appears. This behavior is characterized by periodical oscillations
and enhancement of the response of the system through the
appearance of a stochastic resonance phenomenon in the dynamics of
the two competing species. We then conclude that a stochastic
resonance phenomenon (primary SR) for the interaction parameter
$\beta$ induces another effect of stochastic resonance (secondary
SR) in the dynamics of the species in the presence of colored
noise. We note that for $\tau_c\rightarrow 0$ our results are
consistent with that obtained for the case of white noise (see
Fig.~\ref{snr}). Moreover the coherent time behavior of our
ecosystem and the SR phenomenon are shifted towards higher noise
intensities, in agreement with previous theoretical and
experimantal investigations \cite{Gam,Man,Han}. The presence of a
colored noise assures that the densities of the two species at the
time $t$ depend on the values of the multiplicative noise at the
previous times. This represents a more realistic condition and it
can be useful to explain the time evolution of ecological species,
whose dynamics is strongly affected by the environmental
noise~\cite{Sch,Spa1,Spa2,Laf,Car}.
\section{Acknowledgments}
This work was supported by INTAS Grant 01-450, by INFM and MIUR.

\end{document}